\documentclass[prl,superscriptaddress,tightenlines,twocolumn,balancelastpage]{revtex4}
\usepackage{amsfonts}
\usepackage{amsmath}
\usepackage{amssymb}
\usepackage{graphicx}
\usepackage{epstopdf}
\usepackage{dcolumn}

\begin{document}

\preprint{}
\title[LL in graphite]{Measurement of graphite tight-binding parameters
using high field magneto-reflectance}
\author{L.-C. Tung}
\affiliation{National High Magnetic Field Laboratory, Tallahassee, Florida 32310}
\affiliation{Department of Physics and Astrophysics, University of North Dakota, Grand
Forks, North Dakota 58202}
\author{P. Cadden-Zimansky}
\affiliation{National High Magnetic Field Laboratory, Tallahassee, Florida 32310}
\affiliation{Department of Physics, Columbia University, New York, New York 10027}
\author{J. Qi}
\affiliation{National High Magnetic Field Laboratory, Tallahassee, Florida 32310}
\author{Z. Jiang}
\affiliation{School of Physics, Georgia Institute of Technology, Atlanta, Georgia 30332}
\author{D. Smirnov}
\affiliation{National High Magnetic Field Laboratory, Tallahassee, Florida 32310}
\date{\today }

\begin{abstract}
Magnetic subbands of graphite have been investigated by magneto-infrared
reflectance spectroscopy at 4K in fields up to 31T. Both Schr\"{o}%
dinger-like (K-point) and Dirac-like (H-point) Landau level transitions have
been observed, and their magnetic field dispersion are analyzed by a
newly-derived limiting case of the Slonczewski-Weiss-McClure model. The
values of the band parameters are evaluated without using sophisticated
conductivity peak lineshape analysis. In this work, several less-explored
band parameters are determined from the experimental results and they are
known to result in electron-hole asymmetry and the opening of an energy gap
between the electron and hole bands in multilayer and bilayer graphene
systems.
\end{abstract}

\volumeyear{year}
\volumenumber{number}
\issuenumber{number}
\eid{identifier}
\received[Received text]{date}
\revised[Revised text]{date}
\accepted[Accepted text]{date}
\published[Published text]{date}
\startpage{101}
\endpage{102}
\maketitle

The tight-binding model of graphite's band structure, first calculated by
Wallace\cite{Wal47} and later extended by Slonczewski, Weiss,\cite{Slo58}
and McClure\cite{Mcc57}, has enjoyed a renaissance in recent years due to
its successful application in understanding many properties of single and
few-layer graphene.\cite{Nov04,Cas09} The model, known as
Slonczewski-Weiss-McClure (SWMc) band theory, rests on seven tight binding
parameters, $\gamma _{0}$ to $\gamma _{5}$ and $\Delta $, characterizing
interactions between near neighbor atoms in the lattice. The physical
representations and estimated values of these parameters have been
illustrated and organized by Dresselhaus \textit{et al.}\cite{Dre81} and
Zhang \textit{et al.}\cite{Zha08}. While the SWMc model has been used in
understanding graphitic materials for the last 60 years\cite%
{Dre81,Bra88,Zha08,Mcc60,Ino62,Nak76,Toy77,Doz79,Hub11}, the values of
less-explored band parameters ($\gamma _{4}$, $\gamma _{5}$ and $\Delta $)
remain uncertain. As these parameters may play an expanded role in
interpreting observed asymmetries in the electronics of graphene, there is
an increased incentive to accurately measure them.

Infrared (IR) (magneto)-spectroscopy is a powerful tool in resolving
energies in the band structure of graphitic materials, and has already been
employed to study the unique electronic states of massless Dirac-like
fermion in monolayer graphene\cite{Jia07,Dea07,Hen10}, massive chiral
fermions in bilayer graphene\cite{Zha08,Hen08,Li09,Kuz09}, and rich spectral
features in graphite\cite{Dre81,Toy77,Doz79,Li06,Orl08,Orl09,Chu09}, where
effects attributed to both massless holes and massive electrons have been
seen. The results in graphite have been modeled by assuming that its band
structure is a combination of a monolayer model, where the holes at the
Brillouin zone H-point behave like massless fermions, and an effective
bilayer model (EBM)\cite{Orl08,Orl09,Chu09}, where the electrons at the
K-point behave like massive fermions with an adjusted coupling constant
between layers. However, the EBM ignores the band parameters governing the
interlayer hoppings between unstacked and stacked/unstacked sublattices ($%
\gamma _{3}$, $\gamma _{4}$), next-nearest interlayer hoppings ($\gamma _{2}$%
, $\gamma _{5})$, \ and the on-site energy difference between the stacked
and unstacked sublattices ($\Delta $). Excluding these parameters removes
the possibility of electron-hole ($e-h$) asymmetry, which has been observed
to be significant in bilayer graphene and graphite.\cite%
{Li06,Zha08,Hen08,Chu09,Kuz09}

The interlayer hoppings can also drastically change the band structure near
the charge neutrality point (CNP), i.e. the point where the electron and
hole bands touch, and lift the degeneracy of the doubly degenerate $e-h$
mixed Landau level, (LL-1) and (LL0). Here, we use the level indexing
notation of ref. \cite{Nak76}. The (LL-1) and (LL0) refer to the two LLs
near the CNP, and the term, $hn-$ $em$, refers to a transition from the n-th
hole LL to the m-th electron LL. The splitting of the two $e-h$ mixed LLs
was observed recently by Li \textit{et al.}\cite{Li06}, and the transition
between these two states exhibits a potential 3D to 1D crossover due to
graphite's anisotropic properties.

In this communication, we use IR spectroscopy to measure the magnetic-field
dispersion of graphite's LLs and fit our measurements to a newly-derived
limit of the SWMc model that includes higher-order band parameters. The
intense magnetic fields applied distinguish the two $e-h$ mixed LLs and
split of the $e-h$ symmetry of several interband transitions. These lifted
degeneracies are accounted for by our modified SWMc model, and values for
the set of band parameters are found consistent with those previously
reported.\cite{Dre81,Zha08} 
\begin{figure}[tp]
{{{%
\includegraphics[
natheight=11in,
natwidth=8.5in,
height=3.6in,
width=3.4in
]{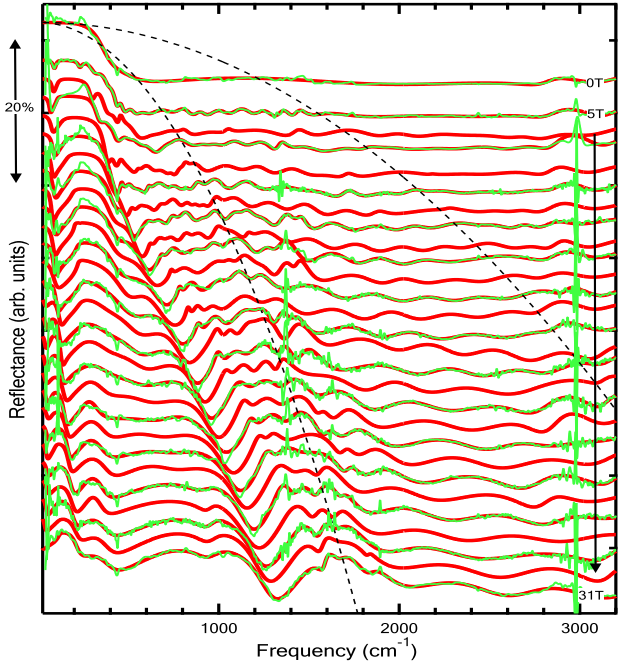}}}}
\caption{Reflectance spectra of graphite: The spectra are vertically shifted
for clarity. The smoothed (original) spectra are shown in thick red(thin
green) lines. Two dashed lines are used as guides for spectral features
following $\protect\sqrt{B}$-dependence. }
\end{figure}

Our measurement set-up uses Kish graphite flakes stabilized on the Scotch
tape and placed in a cryostat held at 4K subjected to magnetic fields of up
to 31T. IR reflectance spectra are measured by a Fourier transform IR
interferometer using light pipe optics. The reflectance is then normalized
by taking the ratio of the sample spectra to that of a gold mirror. The
acquired reflectance spectra are modeled by Drude-Lorentz functions using
RefFit. Sharp or minute features attributable to noise were smoothed prior
to modeling. To show the quality of the modeling, the smoothed reflectance
spectra are plotted in Fig. 1 overlaid with the original spectra at every
other Tesla. Further analysis reveals that the observed oscillatory spectral
features are attached to the K- and H-point transitions, exhibiting linear
in $B$\ and $\sqrt{B}$- dependence, respectively. The features resulting
from $e-h$ mixed LLs are located at low frequencies, while $e-h$ asymmetry
results in splitting in several of the interband transitions at high
frequencies.

The dynamic conductivity, shown in Fig 2, can be calculated from the
modeling of Fig. 1, and exhibits a series of magnetic-field dependent
conductivity peaks. The overall profile of the dynamic conductivity is
consistent with a recent theoretical calculation in the presence of a
moderate magnetic field\cite{Fal11}, and most K-point transitions show
sizeable deviation from the linear dependence. The two lowest-energy modes,
resulting from the (LL-1)-(LL0) and (LL0)-e1 transitions, are displayed in
Fig. 2 (a). It is obvious that magnetic-field dispersion of the (LL0)-e1
transition is not linear, instead exhibiting an interesting crossover from
linear (low-field) to nonlinear (high-field) dependence.

Graphite has been shown to exhibit a universal conductance $%
G_{0}=e^{2}/4\hbar $ for photon energies between $0.1$ to $0.6$eV at zero
magnetic field.\cite{Kuz08} ($1$eV$=8065$cm$^{-1}$) The sheet conductance of
our sample, calculated using $G(\omega )=c_{0}\sigma (\omega )$ with
interlayer distance $c_{0}=3.35\text{\AA}
$, is shown in the inset of Fig. 2 (b). The calculated value is found
larger than $G_{0}=e^{2}/4\hbar $, but closer to $4e^{2}/h$. Accounting for
this discrepancy, one should note that the conductivity is not calculated
from a true Kramer-Kronig transformation, but a result of modeling the
reflectance spectra using finite numbers of Drude-Lorentz functions. As one
cannot precisely assign Lorentz modes outside the measurement range, the
optical weight for the modes located within the measurement range tends to
be overestimated. Moreover, the incident angle and the misalignment (between
the sample and the gold reference) can also affect the result. On the other
hand, the universal conductance is dependent on the short-range disorder,
"ripples" in graphene's atomic structure and the presence of charged
impurities\cite{Che08,Mor08}, all of which are likely to be present in Kish
graphite. The focus of this work is the magnetic-field dispersion of the
LLs, which are not impacted by the frequency-independent universal
conductance.

\begin{figure}[t]
{{{%
\includegraphics[
natheight=8.5in,
natwidth=11in,
height=2.627in,
width=3.4in
]{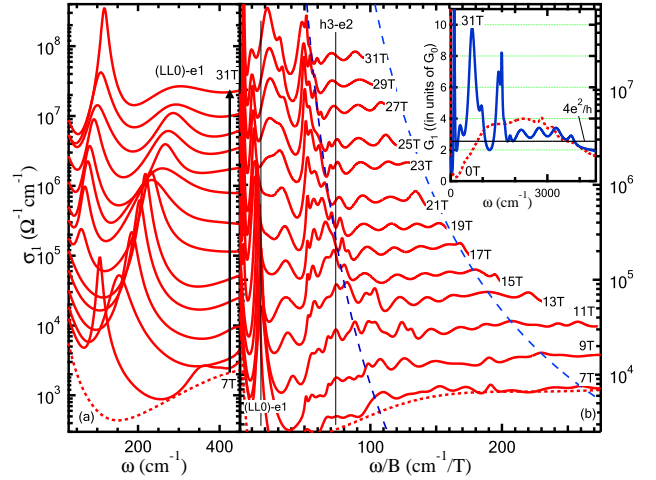}}}}
\caption{Real part ($\protect\sigma _{1}$) of the dynamic conductivity: The
curves are shifted vertically for clarity. For comparison, $\protect\sigma %
_{1}$ at $B=0$T is shown in a dotted line. To help the readers locating the
modes in (b), two vertical lines mark the peaks associated with the (LL0)-e1
and h3-e2 transitions. Two guide lines similar to Fig. 1 follow the H-point
transitions. Inset: Sheet conductance per graphene layer at $B=0$T and $31$T
are shown in units of $G_{0}=e^{2}/4\hbar $.}
\end{figure}

Mode energies extracted from the conductivity peaks of Fig. 2 are plotted in
Fig. 3, with modes following $\sqrt{B}$- dependence shown in (a) and those
following linear dependence shown in (b). The (LL0)-e1 transition follows
linear dependence below $12$T, but $\sqrt{B}$- dependence at higher fields,
and thus is plotted in both panels. Anomalies in mode energies can be found
around the characteristic phonon energies of graphite\cite{Ste04,Fer07},
particularly along several LL transitions (H: -1 to 0, K: (LL0)-e2, h1-e2),
which will be further investigated elsewhere.

In order to invoke the SWMc model, we start with the theoretical observation
that the LL structures for $\gamma _{3}=0$ and $\gamma _{3}\neq 0$ are
similar at high fields\cite{Nak76,Chu09}, and so set $\gamma _{3}\equiv 0$,
which ignores the trigonal warping in the band structure. The energy of the
magnetic subbands can then be obtained using the following self-consistent
Eq.\cite{Mcc60},

\begin{eqnarray}
(n+\frac{1}{2})\widetilde{E}^{2} &=&\varepsilon \lbrack \frac{\varepsilon
(1+\nu ^{2})}{(1-\nu ^{2})^{2}}+\frac{\delta \omega }{(\omega ^{2}-\delta
\omega ^{2})}] \\
&&\pm \sqrt{\lbrack \varepsilon (\frac{2\varepsilon \nu }{(1-\nu ^{2})^{2}}-%
\frac{\omega }{(\omega ^{2}-\delta \omega ^{2})})]^{2}+(\frac{\widetilde{E}%
^{2}}{2})^{2}},  \notag
\end{eqnarray}%
where integer $n$ is the LL index, $\varepsilon =E-E_{3}$, $\widetilde{E}%
^{2}=\tilde{c}^{2}2e\hbar B=qB$ with $\tilde{c}=\frac{\sqrt{3}a_{0}\gamma
_{0}}{2\hbar }$, $\nu =\frac{\gamma _{4}\Gamma }{\gamma _{0}}$ with $\Gamma
=2\cos (\frac{c_{0}k_{z}}{2})$, $\omega =\frac{1}{2}[\frac{(1-\nu )^{2}}{%
E_{1}-E_{3}}-\frac{(1+\nu )^{2}}{E_{2}-E_{3}}]$ and $\delta \omega =\frac{1}{%
2}[\frac{(1-\nu )^{2}}{E_{1}-E_{3}}+\frac{(1+\nu )^{2}}{E_{2}-E_{3}}]$, in
which $c_{0\text{ }}$and $a_{0}$ are lattice parameters. In SWMc model, $%
E_{1}$, $E_{2}$, and $E_{3}$ can be represented as\cite{Mcc60},

\begin{equation}
E_{1,2}=\Delta \pm \gamma _{1}\Gamma +\frac{1}{2}\gamma _{5}\Gamma ^{2}\text{
and }E_{3}=\frac{1}{2}\gamma _{2}\Gamma ^{2}.
\end{equation}%
At H-point, where $\Gamma =0$, the levels are\cite{Mcc60},

\begin{equation}
E=\frac{\Delta }{2}\pm \sqrt{(\frac{\Delta }{2})^{2}+nqB}.
\end{equation}%
Using the modes following $\sqrt{B}$-dependence, we find that the data agree
well with the Fermi velocity $\tilde{c}=1.03\times 10^{6}$ m/s ($\gamma
_{0}=3.18$eV) and assign the two highest energy modes to the $n=-1$ to $n=0$
and $n=-1(-2)$ to $n=2(1)$ transitions. To see if the parameter $\Delta $ is
finite,\ the energy $E_{-1,0}$ for the $n=-1$ to $n=0$ transition is scaled
by a factor of $1+\sqrt{2}$, and plotted in a dotted line in Fig. 3 (a). If $%
\Delta $ is finite, the scaled energy ($1+\sqrt{2})E_{-1,0}$ would be larger
than the interband transition energy $E_{-1,2}$.\cite{Jia07} From the
difference, $\left\vert \Delta \right\vert $ is estimated to be $60$ cm$%
^{-1} $.

For K-point transitions, the levels follows a linear in $B$\ dependence when 
$\tilde{E}$ is small, and can be written as\cite{Mcc60}

\begin{equation}
\varepsilon _{e,h}=[\pm \sqrt{n(n+1)\omega ^{2}+\frac{1}{4}\delta \omega ^{2}%
}-(n+\frac{1}{2})\delta \omega ]qB,
\end{equation}%
in which $\omega $ determines the energy separation between the LLs and $%
\delta \omega $ results in the $e-h$ asymmetry. It is clear from the data
that most K-point modes are not linear at high fields. Though having a
nonlinear magnetic field dispersion, the EBM\cite{Orl09,Chu09} ignores the $%
e-h$ asymmetry; an effect one cannot sufficiently account for by using two
different Fermi velocities\cite{Chu09}, since the correction is apparently
dependent on the LL index $n$. On the other hand, It worths noting that the
EBM\cite{Orl09,Chu09} and the bilayer model\cite{Mcc06} ($\Gamma =1$) can be
obtained by approximating the self-consistent Eq. (1) of the SWMc model as,%
\begin{equation}
\varepsilon ^{2}\cong \frac{2n+1}{2}\widetilde{E}^{2}+\frac{1}{2}\frac{1}{%
\omega ^{2}}-\frac{1}{2}\sqrt{\frac{1}{\omega ^{4}}+4\frac{\varepsilon ^{2}}{%
\omega ^{2}}+\widetilde{E}^{4}}.
\end{equation}%
and use a trial $\varepsilon ^{2}=\frac{2n+1}{2}\widetilde{E}^{2}$ on the
right-hand side.

To derive a working formula to account for the nonlinear magnetic field
dependence of K-point transitions at high fields and with $e-h$ asymmetry,
it is instructive to re-examine the Eq. (1). If the term $(\frac{\widetilde{E%
}^{2}}{2})^{2}$ is neglected, Eq. (1) becomes a simple quadratic equation.
Using graphite's band parameters, the term $(\frac{\widetilde{E}^{2}}{2}%
)^{2} $ can be neglected when $\frac{\varepsilon }{B}>>8$ cm$^{-1}$/T, a
condition can be met for any $n\geq 1$ levels. With this approximation, the
magnetic levels can then be calculated to be

\begin{equation}
\varepsilon _{e,h}=\frac{(1\pm \nu )^{2}}{2}[\mp \frac{1}{\omega \mp \delta
\omega }\pm \sqrt{(\frac{1}{\omega \mp \delta \omega })^{2}+\frac{2(2n+1)}{%
(1\pm \nu )^{2}}qB}].
\end{equation}%
Interestingly, when $\nu =0$, Eq. (6) can be written as

\begin{equation}
\varepsilon _{e,h}=\pm \lbrack \sqrt{(m_{e,h}\widetilde{c}%
^{2})^{2}+2n^{\prime }\hbar \omega _{c}\cdot m_{e,h}\widetilde{c}^{2}}%
-m_{e,h}\widetilde{c}^{2}],
\end{equation}%
where $m_{e,h}\equiv \frac{1}{\widetilde{c}^{2}}\frac{1}{2(\omega \mp \delta
\omega )}$, $n^{\prime }=(n+1/2),$ and $\omega _{c}$ is the cyclotron
frequency. The first term in Eq. (7) agrees with a solution of the Dirac
equation in a magnetic field except for that the value of $n^{\prime }$ was
supposed to be integers.\cite{Sok74} Eq. (7) mimics the relativistic kinetic
energy of a massive particle. Using this analogy, the effect of $\nu $ may
be regarded as a Doppler effect, which causes blueshifts (redshifts) to
electrons (holes). 
\begin{figure}[tp]
{{{%
\includegraphics[
natheight=8.5in,
natwidth=11in,
height=2.627in,
width=3.4in
]{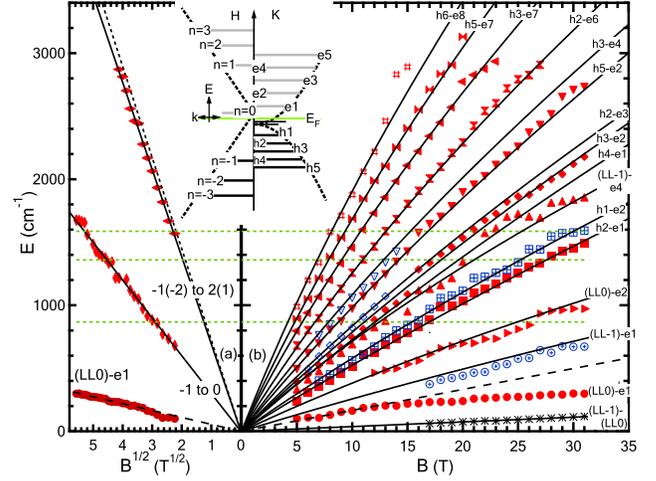}}}}
\caption{Transition energies as a function of magnetic field: Three green
dotted horizontal lines mark the characteristic phonon energies in graphite.
Black solid lines are fittings using Eq. (3) for panel (a) and Eq. (6) for
panel (b). The purpose of the dashed lines is explained in the text. Inset:
Schematic energy diagram at the H- and K-point.}
\end{figure}

In analyzing the $e-h$ asymmetry, potential assignments for the
higher-energy modes can be arbitrary (due to the large number of possible
combinations), thus we use only the lower-energy modes to determine the band
parameters. Good agreement with the data can be achieved using $\omega
q=14.5 $ cm$^{-1}$/T, $\delta \omega q=-3.4$ cm$^{-1}$/T and $\nu =0.05$.
The calculated transition energies are shown in solid lines in Fig. 3 (b)
and the higher-energy modes are assigned to the transitions that best
describe the magnetic-field dispersion of the transition energies. Near the
K-point, i.e. $\gamma _{1}\Gamma >>\Delta +\frac{1}{2}\gamma _{5}\Gamma ^{2}-%
\frac{1}{2}\gamma _{2}\Gamma ^{2}$, the fitting parameters, $\omega $ and $%
\delta \omega $, are related to the band parameters as%
\begin{equation}
\omega \sim \frac{1}{\gamma _{1}\Gamma };\text{ }\delta \omega \sim -\frac{%
\frac{\Delta }{\Gamma ^{2}}+2\gamma _{4}\frac{\gamma _{1}}{\gamma _{0}}+%
\frac{1}{2}\gamma _{5}-\frac{1}{2}\gamma _{2}}{\gamma _{1}{}^{2}}.
\end{equation}%
From the values of fitting parameters, we extract $\gamma _{1}=0.38$eV, $%
\gamma _{4}=0.08$eV and -$\gamma _{2}+\gamma _{5}=0.048$eV. It is generally
agreed\cite{Dre81,Zha08} that $\gamma _{2}=-0.02$eV, so $\gamma _{5}$ is
estimated to be $0.028$eV.

For the (LL-1) and (LL0) levels (i.e. $n=0$), one can solve the cubic Eq. in
Ref. \cite{Ino62} to find

\begin{equation}
\varepsilon =0\text{ and }\varepsilon \sim -\delta \omega qB,
\end{equation}%
respectively. Since the (LL0) level will be partially occupied by electrons,
the transition between the (LL-1) and (LL0) levels are made by those
electrons away from the K-point. The parameter $\delta \omega (k_{z}\sim
k_{F})q$ is found to be $-3.7$cm$^{-1}$/T, which is slightly larger than $%
\delta \omega (k_{z}=0)q=-3.4$cm$^{-1}$/T; therefore, the sublattice energy
difference $\Delta $ is likely to be positive.

The (LL0)-e1 transition exhibits an unusual crossover. At lower fields, its
energy agrees with the prediction of the linear SWMc model with a slope of $%
17$cm$^{-1}$/T. (dashed line in Fig. 3 (b)) At higher fields, the peak
broadens quickly with increasing magnetic field, and the magnetic-field
dispersion becomes Dirac-like with a slope of $54$cm$^{-1}$/T$^{1/2}$.
(dashed line in Fig. 3 (a)) A similar behavior has been reported previously%
\cite{Li06}, where the (LL0)-e1 mode has a larger slope for $B\leq 8.5$T,
but a smaller slope for $B\geq 8.5$T. In addition, we find that $%
E_{(LL-1)-(LL0)}+E_{(LL0)-e1}<E_{(LL-1)-e1}$, a discrepancy that cannot be
accounted for by simple pictures assuming LLs are independent with each
other. Except for the (LL0)-e1 transition, all other transitions are well
described by the modified SWMc model (Eq. (6)), so the underlying mechanism
for this anomaly is specific to the (LL0) level. Anomalous phenomena have
been observed pertaining to this level, and it has been conjectured that a
charge density wave\cite{Yos81,Iye82,Sug84} has an effect specific to it,
though it should be noted that these anomalies were found at lower
temperatures and at higher fields.

In summary, magnetic subbands in Kish graphite are investigated using
magneto-infrared reflectance spectroscopy up to 31T. The high magnetic field
distinguishes the two $e-h$ mixing LLs near the CNP and the splitting of the
interband transitions. We propose a new limit of the SWMc model to describe
the K-point transitions and use it to estimate the values of the band
parameters, 
\begin{eqnarray*}
\gamma _{0} &=&3.18\text{eV, }\gamma _{1}=0.38\text{eV, }\gamma _{2}=-0.02%
\text{eV, }\gamma _{3}\equiv 0\text{, } \\
\gamma _{4} &=&0.08\text{eV, }\gamma _{5}=0.028\text{eV, and }\left\vert
\Delta \right\vert =0.008\text{eV.}
\end{eqnarray*}

\begin{acknowledgments}
This work was supported by the DOE (DE-FG02-07ER46451) and the NSF
(DMR-0820382). The IR measurement was carried out at the National High
Magnetic Field Laboratory, which is supported by NSF Cooperative Agreement
No. DMR-0654118, by the State of Florida, and by the DOE.
\end{acknowledgments}

\end{document}